
\newif\iffigs\figstrue

%
\let\useblackboard=\iftrue
%
%
\newfam\black

\input harvmac.tex


\def\Title#1#2{\rightline{#1}
\ifx\answ\bigans\nopagenumbers\pageno0\vskip1in%
\baselineskip 15pt plus 1pt minus 1pt
\else
\def\listrefs {
\footatend\vskip 1in\immediate\closeout\rfile\writestoppt
\baselineskip=14pt
\centerline{{ \bf References}}
\bigskip{\frenchspacing%
\parindent=20pt\escapechar=` \input
refs.tmp\vfill\eject}\nonfrenchspacing}
\pageno1\vskip.8in\fi \centerline{\titlefont #2}\vskip .5in}

\ifx\answ\bigans\def\tcbreak#1{}
\else\def\tcbreak#1{\cr&{#1}}\fi
\useblackboard
\message{If you do not have msbm (blackboard bold) fonts,}
\message{change the option at the top of the tex file.}
\font\blackboard=msbm10 
\font\blackboards=msbm7
\font\blackboardss=msbm5
\textfont\black=\blackboard
\scriptfont\black=\blackboards
\scriptscriptfont\black=\blackboardss
\def\Bbb#1{{\fam\black\relax#1}}
\else
\def\Bbb#1{{\bf #1}}
\fi
\def\yboxit#1#2{\vbox{\hrule height #1 \hbox{\vrule width #1
\vbox{#2}\vrule width #1 }\hrule height #1 }}
\def\fillbox#1{\hbox to #1{\vbox to #1{\vfil}\hfil}}
\def\ybox{{\lower 1.3pt \yboxit{0.4pt}{\fillbox{8pt}}\hskip-0.2pt}}
\def\np#1#2#3{Nucl. Phys. {\bf B#1} (#2) #3}
\def\pl#1#2#3{Phys. Lett. {\bf #1B} (#2) #3}

\def\physrev#1#2#3{Phys. Rev. {\bf D#1} (#2) #3}

\def\cmp#1#2#3{Comm. Math. Phys. {\bf #1} (#2) #3}

\def\jhep#1#2#3{{{JHEP.{\bf {#1}}({#2}){#3}}}}
\def\atmp#1#2#3{Adv. Theor. Math. Phys. {\bf #1}(#2) #3}

\def\comments#1{}

\def\half{{1\over 2}}

\def\a{\alpha}

\def\II{\relax{I\kern-.07em I}}

\def\IZ{\relax\ifmmode\mathchoice
{\hbox{\cmss Z\kern-.4em Z}}{\hbox{\cmss Z\kern-.4em Z}}
{\lower.9pt\hbox{\cmsss Z\kern-.4em Z}}
{\lower1.2pt\hbox{\cmsss Z\kern-.4em Z}}\else{\cmss Z\kern-.4em
Z}\fi}
\def\IB{\relax{\rm I\kern-.18em B}}

\def\ID{\relax{\rm I\kern-.18em D}}
\def\IE{\relax{\rm I\kern-.18em E}}
\def\IF{\relax{\rm I\kern-.18em F}}
\def\IG{\relax\hbox{$\inbar\kern-.3em{\rm G}$}}
\def\IGa{\relax\hbox{${\rm I}\kern-.18em\Gamma$}}
\def\IH{\relax{\rm I\kern-.18em H}}
\def\II{\relax{\rm I\kern-.18em I}}
\def\IK{\relax{\rm I\kern-.18em K}}
\def\IP{\relax{\rm I\kern-.18em P}}

\useblackboard
\def\IZ{\relax\Bbb{Z}}
\fi

\font\cmss=cmss10 \font\cmsss=cmss10 at 7pt
\def\IR{\relax{\rm I\kern-.18em R}}

\def\BR{\IR}
\def\BZ{\IZ}
\def\BR{\IR}

\def\tilde{\widetilde}

%
%

\def\lim{{lim}}


\def\SUSY#1{{{\cal N}= {#1}}}                   

\def\wdg{{\wedge}}                              
\def\MR#1{{{\BR}^{#1}}}               
\def\MR#1{{{\BR}^{#1}}}               
\def\MS#1{{{\bf S}^{#1}}}               
\def\MT#1{{{\bf T}^{#1}}}               
\def\MHT#1{{{\bf \widetilde{T}}^{#1}}}               

\def\px#1{{\partial_{#1}}}              





\def\rep#1{{{\bf {#1}}}}                      



\def\hepth#1{{\it hep-th/{#1}}}

\def\frac#1#2{{{{#1}}\over {{#2}}}}           


\def\u{{\mu}}
\def\v{{\nu}}
\def\b{{\beta}}

\def\lam{{\lambda}}





\def\tw{{\alpha}}         
\def\etw{{\eta}}         

\def\bZ{{\overline{Z}}}
\def\bW{{\overline{W}}}
\def\qx#1{{\partial^{{#1}}}}

\def\ve{{\hat{e}}}
\def\TNS{{\rho}}        
\def\TBNT{{{\bf TN}}}   

\def\Ur#1{U_{\left[#1\right]}(|\vec{r}|)}
\def\tp{\tilde{\psi}}

\def\vr{|\vec{r}|}

%
%
%
%
\Title{hep-th/9812172, PUPT-1826,  ITEP-TH-68/98}
{\vbox{\centerline{Instantons on a Non-commutative $T^4$}
\bigskip
\centerline{from}
\bigskip
\centerline{Twisted (2,0) and Little String Theories}
}}
\bigskip
\centerline{
Yeuk-Kwan E. Cheung\footnote{$^1$}{cheung@viper.princeton.edu},
Ori J. Ganor\footnote{$^2$} {origa@puhep1.princeton.edu},
Morten Krogh\footnote{$^3$}{krogh@phoenix.princeton.edu} and
Andrei Yu. Mikhailov\footnote{$^4$} {andrei@puhep1.princeton.edu}
\footnote{$^*$}{On leave from the Institute of Theoretical
and Experimental Physics, Moscow, Russia.}
}
\medskip
\centerline{Department of Physics}
\centerline{Princeton University}
\centerline{Princeton, NJ 08544-0708, USA}
\bigskip
\noindent
We show that the moduli space of the $(2,0)$ and little-string
theories compactified on $T^3$ with R-symmetry twists is equal
to the moduli space of $U(1)$ instantons
on a non-commutative $T^4$. The moduli space of $U(q)$ instantons
on a non-commutative $T^4$ is obtained from little-string theories
of NS5-branes at $A_{q-1}$ singularities with twists.
A large class of gauge theories with ${\cal N}=4$
SUSY in 2+1D and ${\cal N}=2$ SUSY in 3+1D are limiting
cases of these theories. Hence, the moduli spaces of these gauge
theories can be read off from the moduli spaces of
instantons on non-commutative tori.
We study the phase
transitions in these theories and the action of T-duality.
On the purely mathematical side, we give a prediction for
the moduli space of 2 $U(1)$ instantons on a non-commutative $T^4$.

\Date{December, 1998}

\nref\rSWI{N. Seiberg and E. Witten,
  {``Electric-Magnetic Duality, Monopole Condensation,
  and Confinement in $N=2$ Supersymmetric Yang-Mills Theory,''}
  \np{426}{1994}{19}, \hepth{9407087}.}

\nref\rKV{S. Kachru and C. Vafa,
  {``Exact Results For $N=2$ Compactifications
  Of Heterotic Strings,''} \np{450}{1995}{69}, \hepth{9505105}.}

\nref\rWitCOM{E. Witten,
  {``Some Comments on String Dynamics,''}
  \hepth{9507121},
  published in {\it ``Future perspectives in string theory,''} 501-523.}

\nref\rSeiVBR{N. Seiberg,
  {``New Theories in Six-Dimensions and
  Matrix Description of M-theory on $T^5$ and $T^5/Z_2$,''}
  \hepth{9705221}, \pl{408}{1997}{98}}

\nref\rDVVQ{R. Dijkgraaf, E. Verlinde, H. Verlinde,       
 {``BPS Quantization Of The 5-Brane,''}
  \np{486}{1997}{77}, \hepth{9604055}}
\nref\rDVVS{R. Dijkgraaf, E. Verlinde and H. Verlinde,
  {``BPS Spectrum Of The 5-Brane And Black-Hole Entropy,''}
  \np{486}{1997}{89}, \hepth{9604055}}

\nref\rWitFBR{E. Witten,
  {``Solutions Of Four-Dimensional Field Theories Via M Theory,''}
  \np{500}{1997}{3--42}, \hepth{9703166}}

\nref\rCGK{Y.-K.E. Cheung, O.J. Ganor and M. Krogh,
  {``On the twisted $(2,0)$ and Little-String Theories,''}
  \np{536}{1998}{175}, \hepth{9805045}}

\nref\rBluInt{J.D. Blum and K. Intriligator,
  {``New Phases of String Theory and 6d RG Fixed Points via Branes at
  Orbifold Singularities,''} \np{506}{1997}{199}, \hepth{9705044}}

\nref\rIntNEW{K. Intriligator,
  {``New String Theories in Six Dimensions via Branes at Orbifold
  Singularities,''} \atmp{1}{1998}{271}, \hepth{9708117}}

\nref\rDM{M.R. Douglas and G. Moore, 
         {``D-branes, Quivers, and ALE Instantons,''}
         \hepth{9603167}}

\nref\rGanSet{O.J. Ganor and S. Sethi,
   {``New Perspectives On Yang-Mills Theories With 16 Supersymmetries,''}
     \jhep{01}{1998}{007}, \hepth{9712071}}

\nref\rKapSet{A. Kapustin and S. Sethi,
  {``The Higgs Branch of Impurity Theories,''}
  \atmp{2}{1998}{571}, \hepth{9804027}}

\nref\rWitNGT{E. Witten,
  {``New ``Gauge'' Theories In Six Dimensions,''}
  \atmp{2}{1998}{61}, \hepth{9710065}}

\nref\rBDS{T. Banks, M.R. Douglas and N. Seiberg,
  {``Probing F-theory With Branes,''}
  \pl{387}{1996}{278--281}, \hepth{9605199}}

\nref\rSeiIRD{N. Seiberg,
  {``IR Dynamics on Branes and Space-Time Geometry,''}
  \pl{384}{1996}{81--85}, \hepth{9606017}}

\nref\rSWGDC{N. Seiberg and E. Witten,
  {``Gauge Dynamics And Compactification To Three Dimensions,''}
  \hepth{9607163}}

\nref\rDouglas{ M.R.~Douglas, ``Branes within Branes'', \hepth{9512077}}

\nref\rCDS{A. Connes, M.R. Douglas and A. Schwarz,
  {``Noncommutative Geometry and Matrix Theory: Compactification on Tori,''}
  \jhep{02}{1998}{003}, \hepth{9711162}}

\nref\rDH{M.R. Douglas and C. Hull,
  {``D-branes and the Noncommutative Torus,''}\break
  \jhep{02}{1998}{008}, \hepth{9711165}}

\nref\rANS{A. Astashkevich, N. Nekrasov and A. Schwarz,
     {``On noncommutative Nahm transform,''}
   \hepth{9810147}}

\nref\rHoWuWu{P.-M. Ho, Y.-Y. Wu and Y.-S. Wu, 
    {``Towards a Noncommutative Geometric Approach to 
       Matrix Compactification,''}
    \physrev{58}{1998}{026006}, \hepth{9712201}} 

\nref\rHoWu{P.-M. Ho, Y.-S. Wu,
{``Noncommutative Gauge Theories in Matrix Theory,''}
  \physrev{58}{1998}{066003}, \hepth{9801147}}

\nref\rNekSch{N. Nekrasov and A. Schwarz,
  {``Instantons on noncommutative $R^4$ and (2,0) superconformal       
  six dimensional theory,''} \cmp{198}{1998}{689}, \hepth{9802068}}

\nref\rMicha{M. Berkooz,
  {``Non-local Field Theories and the Non-commutative Torus,''}
  \pl{430}{1998}{237}, \hepth{9802069}}

\nref\rBig{D. Bigatti, {``Non commutative geometry for outsiders,''}
  \hepth{9802129}} 

\nref\rCK{Y.-K. E. Cheung and M. Krogh,
  {``Noncommutative Geometry from 0-branes in a Background B-field,''}
  \np{528}{1998}{185-196},    \hepth{9803031}}

\nref\rKO{T. Kawano and K. Okuyama,
  {``Matrix Theory on Noncommutative Torus,'' }
 \pl{433}{1998}{29}, \hepth{9803044}} 

\nref\rAAS{ F. Ardalan, H. Arfaei and M.M. Sheikh-Jabbari,
{``Mixed Branes and M(atrix) Theory on Noncommutative Torus,''}
 \hepth{9803067}}

\nref\rHo{P.-M. Ho,
   {``Twisted Bundle on Quantum Torus and
     BPS States in Matrix Theory,''}
  \pl{434}{1998}{41}, \hepth{9803166}}

\nref\rMZ{B. Morariu, B. Zumino,
  {``Super Yang-Mills on the Noncommutative Torus,''}
  \pl{433}{1998}{279}, \hepth{9807198}}

\nref\rAASJ {F. Ardalan, H. Arfaei and M.M. Sheikh-Jabbari,
  {``Noncommutative Geometry form Strings and Branes,''}
 \hepth{9810072}}

\nref\rHVer{C. Hofman and E. Verlinde,
    {``U-duality of Born-Infeld on the Noncommutative  Two Torus,''}
 \hepth{9810116}}

\nref\rWati{W. Taylor,
  {``D-brane field theory on compact spaces,''}
  \pl{394}{1997}{283}, \hepth{9611042}}


\nref\rNek{N. Nekrasov,
  {``Five-Dimensional Gauge Theories
   and Relativistic Integrable Systems,''}
  \np{531}{1998}{323}, \hepth{9609219}}

\nref\rFMW{R. Friedman, J. Morgan, and E. Witten,
  {``Vector Bundles And F Theory,''}
  \cmp{187}{1997}{679}, \hepth{9701162}}

\nref\rBJPS{M. Bershadsky, A. Johansen, T. Pantev and V. Sadov,
  {``Four-Dimensional Compactifications of F-theory,''}
  \np{505}{1997}{165}, \hepth{9701165}}

\nref\rGriHar{Griffiths and Harris,
   {\it ``Principles of Algebraic Geometry''},
   Wiley-Interscience, New-York, 1978}

\nref\rABS{O. Aharony, M. Berkooz and N. Seiberg,
  {``Linear Dilatons, NS Five-Branes and Holography,''}
  \jhep{9810}{1998}{004}, \hepth{9808149}}
    
\newsec{Introduction}
In recent years, starting with the work of \rSWI,
 the moduli-spaces of vacua have been found
for a large class of gauge theories with 8 super-charges in 3+1D
and in 2+1D. 
 These solutions were derived from string dualities in \rKV\ and
the works that followed.
String theory also suggested the existence of new theories 
in six dimensions \refs{\rWitCOM,\rSeiVBR}
(see also \refs{\rDVVQ-\rDVVS}).
Compactification of these theories to 3+1D reduces, in certain limits
of the external parameter spaces, to ordinary gauge theories.


In this paper we will study compactifications of certain 6 dimensional 
theories down to $3$ dimensions and examine their low energy behaviour. 
As we will see, all the previously solved gauge theories with $\SUSY{2}$
supersymmetry and
$SU(N_1)\times\cdots\times  SU(N_r)$ gauge groups \rWitFBR\ can be
recovered at special limits of the external parameters of the 
compactification.
We will start with the 6 dimensional theories that
are the world-volume theories on $k$ NS5-branes 
in type-IIA in the limit of vanishing string coupling keeping the string 
tension fixed \rSeiVBR. We denote this theory $S_A(k)$. It has 
$(2,0)$ supersymmetry. There is a similar theory coming from $k$
NS5-branes in type-IIB in the limit of vanishing string coupling 
keeping the string tension fixed.
We denote this theory $S_B(k)$. It has $(1,1)$ supersymmetry.
$S_A(k)$ and $S_B(k)$ are often referred to as the little-string theories.
They both have an inherent scale, $m_s$. In the limit $m_s \rightarrow 
\infty$, $S_A(k)$ becomes the theory on the world-volume of $k$ M5-branes
 -- the so called $(2,0)$ theory.

We will compactify these theories down to 3 dimensions. These theories        
have 16 super-charges, so if they are compactified on $\MT{3}$ the resulting 
theories will have $\SUSY{8}$ supersymmetry in three dimensions.
The low energy behavior of $\SUSY{8}$ theories is trivial.
Instead we want to study theories 
with $\SUSY{4}$ supersymmetry, i.e. 8 super-charges.
So we have to compactify 
in a way that breaks half the supersymmetry. We will do 
that as in \rCGK\ by introducing
 holonomies of the R-symmetry around the three circles in 
$\MT{3}$. To preserve half of the supersymmetries the holonomies were 
chosen inside a $SU(2)$ subgroup of the $Spin(4)$ R-symmetry group. 
The low energy behaviour of a $\SUSY{4}$ theory in $D=3$ is a 
sigma-model with the moduli-space of vacua as the target-space. 
So the low energy behaviour is given by the moduli-space 
of vacua and the its metric.

In \rCGK\ the case of $k=2$ was studied in detail. 
The moduli space of vacua was identified explicitly. 
 For general $k$ it was conjectured that the moduli space
 of vacua was given as a moduli space of instantons on 
non-commutative $\MT{4}$. The purpose of this paper will be to 
prove and generalize this and, at the same time, make the claim more 
precise.
 
Let us start by identifying all the parameters of the 
compactification. Consider $S_A(k)$ compactified on $\MT{3}$. 
The scale of $S_A(k)$ is $m_s$, the string mass. The $\MT{3}$ 
is specified by a metric. For simplicity we will take it to 
be rectangular. It is easy to incorporate the more 
general case. Furthermore there can be a flux of the 2-form 
$B^{NS}$ field of type IIA through 2-cycles in the $\MT{3}$. 
 For simplicity we set $B^{NS}=0$. It is again not hard to incorporate
the more general case. Now we come to the most interesting 
parameters -- the twists. The R-symmetry group of $S_A(k)$ is 
$Spin(4)_R$, corresponding to transverse rotations. The twists are 
taken inside 
\eqn\introtwist{
U(1)_R \subset SU(2)_B \subset SU(2)_B \times SU(2)_U = 
Spin(4)_R 
}
This preserves 8 of the 16 super-charges. There is a twist, 
$\alpha_i$, along each of the 3 circles. The $\alpha_i$'s are 
periodic 
\eqn\periode{
\alpha_i \rightarrow \alpha_i +2 \pi,\qquad  i=1,2,3
}       
The twists can be described in the following way. States that are charged
under $U(1)_R$ receive a phase shift in traversing a circle.
In other words, momentum along the circle is shifted 
from $n \over R$ to ${n - {\alpha \over 2\pi} }\over R$.
By performing T-duality along all circles of the $\MT{3}$ 
we get $S_B(k)$ on another $\MT{3}$. Momentum has been exchanged 
with winding, so the T-dual of the twists has the following 
description. States that are charged under $U(1)_R$ have fractional 
winding numbers; $n - \alpha \over 2\pi$ instead of n. 
We call this kind of twist an ``$\eta$-twist.'' By combining these 
two types of twists we learn that the most general twist around 
a circle shifts both momentum and winding. In other words the 
$S_A(k)$ compactification on $\MT{3}$ depends on 6 parameters
\eqn\sekspar{
\alpha_i,\eta_i,\qquad    i=1,2,3,
}
where $\alpha_i$ shifts momentum and $\eta_i$ shifts winding. 
 The $\alpha_i$'s have a clear geometrical interpretation. In 
traversing the circle the transverse space is rotated. 
The $\eta_i$'s are harder to visualize. They are geometrical in the 
T-dual $S_B(k)$.       

We can actually generalize this system even more. Instead of 
$k$ NS5-branes we can consider $k$ NS5-branes on top of an
$A_{q-1}$ singularity. In other words the transverse space to the 
NS5-branes is $\MR{4} / Z_q$, where $Z_q$ is a subgroup of 
$U(1)_R$. $U(1)_R$ is still a symmetry of this space, so we can 
twist as before. These theories have 8 super-charges in 
6 dimensions. The $U(1)_R$ is a global symmetry
which commutes with super-charges.
The twists, therefore, do not break any more supersymmetry, so 
the compactified theory still has $\SUSY{4}$ in 3 dimensions.

Theories of branes on top of an ADE singularity have been 
studied in \refs{\rBluInt,\rIntNEW}. These 6 dimensional theories are,
loosely speaking, quiver gauge theories \rDM\ coupled 
to tensor theories or vice versa, depending on 
whether it is in type-IIA or type-IIB.

The 3 dimensional theory, obtained after compactification 
with twists, has a low energy description as a sigma-model
with a target-space, which is equal to the moduli-space 
of vacua. In this paper we will prove that the moduli-space 
of vacua is equal to the moduli-space of $k$ $U(q)$ instantons 
on a non-commutative $\MT{4}$. The non-commutativity is set        
by the 6 parameters $\alpha_i$ and $\eta_i$.
This generalizes the case of compactification without twists
where the moduli-space of the theories
turns out to be the moduli-space of ordinary instantons
\refs{\rIntNEW,\rGanSet}.

This result implies similar results for all the theories which 
are special cases of this. This includes firstly the $(2,0)$
theory which can be obtained from $S_A(k)$ by $m_s \rightarrow 
\infty$. Secondly, it includes all three-dimensional
$U(k)$ gauge theories with adjoint matter. 
By incorporating the $A_q$ singularity it 
also includes all gauge theories with group $U(k) \times 
\cdots \times U(k)$ and matter in $(k, \bar{k}, 1, \ldots ,1) + 
permutations$. By taking the gauge coupling
 to zero in some $U(k)$ we can get theories with the gauge group 
 being $U(k) \times \cdots \times U(k)$ with fundamental and bi-fundamental 
matter in various combinations with generic masses.

Our results imply that all these 2+1D gauge
theories have a moduli-space 
of vacua equal to the moduli-space of vacua of instantons on 
non-commutative $\MR{3} \times \MS{1}$. In the case of mass 
deformed $\SUSY{8}$ this result was derived earlier in \rKapSet.

By decompactifying one 
circle similar results hold for the moduli space of 
4 dimensional gauge theories on $\MR{3} \times \MS{1}$.       

We also find that for certain discrete values of the 
twists there are Higgs branches emanating from some locus of the 
Coulomb branch. We will identify these and calculate their 
dimensions. We will also calculate the existence of these 
branches from pure field theoretic arguments and find agreement 
in the structure of the Higgs branches. These branches 
generalize a branch found in \rCGK.

Moreover,
combining our results with the formulas in \rCGK\ for the special
case of $q=1$ and $k=2$, we get a prediction for the moduli-space
of two $U(1)$ instantons on a non-commutative $\MT{4}$.
This is a $K3$ (projecting out the center of mass)
and the exact point in moduli-space was given
in \rCGK\ as a function of the twists.

The organization of the paper is as follows. In section (2)
we present the proof that the moduli space is equal to the 
moduli space of instantons on non-commutative $\MT{4}$. 
In section (3) we have a short review of non-commutative gauge 
theories. In section (4) we use this information about 
non-commutative gauge theory to make the claim about the 
moduli-space of non-commutative instantons precise and 
discuss some features of it. In section (5) we
describe the decompactification limit to 3+1D (compactification
of the 5+1D theories on $\MT{2}$ with twists).
In section (6) we present a more detailed geometrical formulation
of $\tw$-twists and especially $\etw$-twists.
We conclude with a summary of the results and possible further
direction.
\newsec{The solution}
\def\Step#1{\vskip 0.5cm {\it Step {#1}:}}

In this section, we derive the solution to the moduli space
of the twisted theory.
To construct the solution we will start with type-IIA on a space
$$
\MR{2,1}\times\MT{3}\times_{\vec{\tw}}\MR{4},
$$
where $\times_{\vec{\tw}}$ means that locally the space looks like
$\MR{2,1}\times\MT{3}\times\MR{4}$ but as we go around a cycle
of the $\MT{3}$ we have to twist the transverse space $\MR{4}$
by the appropriate element of $Spin(4)$ corresponding to the twist.
Now we take $k$ NS5-branes and let them stretch along
$\MR{2,1}\times\MT{3}$ and the origin of $\MR{4}$.
The question what is the low-energy effective action for this
system in the limit that the type-IIA string
coupling constant $\lam\rightarrow 0$.

As will be clear later on, it is easier to solve the problem if
we first replace the transverse $\MR{4}$ with another manifold $M_4$.
In the limit that the curvature of $M_4$ is small at the position
of the NS5-branes the switch from $\MR{4}$ to $M_4$ will not make
a big difference.
Moreover, we can argue that the quantum fluctuations
in the transverse position of the
NS5-brane are related to the fluctuations  of the scalars of $S_A(k)$ as,
$$
x\sim m_s^{-3}\lam\Phi,
$$
and for energy scales $m_s$, $\Phi$ is of the order of $m_s^2$.
In the limit $\lam\rightarrow 0$, the transverse fluctuations of
the NS5-brane go to zero and if the point in $M_4$ is smooth, it 
would seem that the dynamics of the NS5-brane will be the same
as on $\MR{4}$.
This argument should be taken with caution since the actual solitonic
solution of the NS5-brane has a cross-section of about $m_s$.
In any case, we will not have to rely on this argument.

The manifold $M_4$ that we will use is the Taub-NUT space.
The metric is,
\eqn\tnmet{
ds^2 = \TNS^2 U(dy - A_i dx^i)^2 + U^{-1} (d\vec{x})^2,\qquad
i=1\dots 3,\qquad 0\le y\le 2\pi.
}
where,
$$
U = \left(1 + {{\TNS}\over {2|\vec{x}|}}\right)^{-1},
$$
and $A_i$ is the gauge field of a monopole centered at the origin.

The Taub-NUT space has the following desirable properties
(these properties were also used in \rWitNGT),
\item{(1)}
If we excise the origin, what remains is a circle fibration
over $\MR{3}-\{0\}$. Eqn\tnmet\ is written such that
$\vec{x}$ is the coordinate on this base $\MR{3}-\{0\}$.
 For $|\vec{x}|$ restricted to a constant, the fibration is
exactly the Hopf fibration of $\MS{3}$ over $\MS{2}$.

\item{(2)}
The origin $\vec{x} = 0$ is a smooth point.

\item{(3)}
As $|\vec{x}|\rightarrow\infty$ the radius of the fiber
becomes $\TNS$.

\item{(4)}
The space has a $U(1)$ isometry group that preserves the
origin $\vec{x}=0$. An element $g(\theta) = e^{i\theta}\in U(1)$
acts by $y\rightarrow y+\theta$.
It also acts on the tangent space $\MR{4}$
at the origin by embedding $e^{i\theta}$ inside
$$
U(1)\rightarrow SU(2)_L\rightarrow
(SU(2)_L\otimes SU(2)_R)/\BZ_2 = SO(4).
$$

Now that we have replaced the transverse $\MR{4}$ with a Taub-NUT space
we have $k$ NS5-branes on the space,
$$
\MR{2,1}\times\MT{3}\times_{\vec{\tw}} \TBNT(\rho).
$$
The $\tw$-twists are incorporated as follows.
As we go around a cycle of $\MT{3}$ we have to act on the fiber
$\TBNT(\rho)$ with $g(\tw_i)$ where $\tw_i$ is the appropriate
twist.
 In the limit $\rho\rightarrow\infty$,
$\TBNT(\rho)$ becomes $\MR{4}$
and the isometry $g(\tw_i)$ becomes the element in $SO(4)$ that
we have used for the twist.
The virtue of working with $\TBNT(\rho)$ instead of $\MR{4}$
is that at $\vec{x}=\infty$ the circle fiber becomes of
finite size which will help in subsequent dualities.

To generalize the construction to the case of $k$ NS5-branes
at an $A_{q-1}$ singularity, $\MR{4}/\BZ_q$,
we replace the transverse $\MR{4}/\BZ_q$ with a $q$-centered
Taub-NUT space, $\TBNT_q(\rho)$ with radius $\rho\rightarrow\infty$.
This space has similar properties,
\item{(1')}
If we excise the origin, what remains is a circle fibration
over $\MR{3}-\{0\}$.
 For $|\vec{x}|$ restricted to a constant, the fibration is
is a circle bundle over $\MS{2}$ with first Chern-class $c_1 = q$.

\item{(2')}
Near the origin $\vec{x} = 0$, $\TBNT_q$ looks like
$\MR{4}/\BZ_q$.

\item{(3')}
As $|\vec{x}|\rightarrow\infty$ the radius of the fiber
becomes $\TNS$.

\item{(4')}
The space has a $U(1)$ isometry group that preserves the
origin $\vec{x}=0$. An element $g(\theta) = e^{i\theta}\in U(1)$
 acts  at $\vec{x}=\infty$ by $y\rightarrow y+\theta$.
It also acts on the tangent space $\MR{4}/\BZ_q$
at the origin by embedding $e^{i\theta}$ inside
$$
U(1)\rightarrow SU(2)_L\rightarrow
(SU(2)_L\otimes SU(2)_R)/\BZ_2 = SO(4).
$$
Note that the discrete $\BZ_q$ by which we mod out is a subgroup
of the same $U(1)\subset SU(2)_L$ as well.

\subsec{Chains of Dualities}

We have seen that the twisted compactified little-string theories
can be realized as follows. Start with type-IIA on
$\MR{2,1}\times \MT{3}\times \TBNT_q$,
where the radii of $\MT{3}$ are $R_i$ (of the order of $m_s$)
and the radius of the fiber of
the Taub-NUT space is taken to be $\TNS$.
Put $k$ NS5-branes on $\MR{2,1}\times\MT{3}$ and study the limit,
$$
\lam\rightarrow 0,\qquad m_s\TNS\rightarrow\infty.
$$
In principle, we could probably settle on a constant $m_s\TNS$ as well,
since the transverse fluctuations of the NS5-brane are small.
However, the transverse size of the NS5-brane,
as a solitonic object, is of the order of $m_s^{-1}$.
Therefore, to be on the safe side, we take $m_s\TNS\rightarrow\infty$.
The technique for solving theories with 8 supersymmetries is \rKV\
to identify a parameter that decouples from the vector-multiplet
and such that at one limit of this parameter the theory
is described by gauge theory (or little-string theory, in our
case) and in another limit a dual description becomes weakly
coupled. In that second limit, the theory is no longer described
by the gauge theory but the vacuum structure remains the same
and is determined by the classical equations of motion.
This method was also applied in \refs{\rBDS,\rSeiIRD,\rWitFBR}.

In our case, to solve the problem we take the limit of strong
coupling keeping the Taub-NUT radius large.
\eqn\gvul{
\lam\rightarrow\infty,\qquad m_s\TNS\rightarrow \infty,
}
We will also require that $\lam (m_s\TNS)^{-3}\rightarrow\infty$.
We can think of $\TNS$ as being fixed but very large and
$\lam\rightarrow\infty$ much faster.
We will not show that this corresponds to a parameter that
is in a hyper-multiplet (and hence decouples from the vector-multiplets)
but this is the basic assumption.
Recall that in 2+1D hyper-multiplets and vector-multiplets
can be distinguished with the help of the
$U(1)_R\otimes SU(2)_U$ symmetry which is the unbroken subgroup
of \introtwist. The scalar fields of a  vector-multiplet are invariant
under $SU(2)_U$ while the scalar fields of a hyper-multiplet are in
the $\rep{2}$ (see \rSWGDC).
(The dilaton, which is a singlet,
is a quadratic expression in these fields.)
Similarly, the fermions of a hyper-multiplet 
are invariant under $SU(2)_U$
and the fermions of a vector-multiplet
are in the  $\rep{2}$.

The next step is to use string-dualities to convert the region
\gvul\ to a weakly coupled theory.

At this point we have 
$k$ NS5-branes in 
type-IIA on $\MR{2,1}\times \MT{3}\times \TBNT_q$
with string coupling $\lam$, string scale $m_s$,
$\MT{3}$-radii $R_i$, and twists $\tw_i$.
 For simplicity, we assume that $\MT{3}$ is of the form
$\MS{1}\times\MS{1}\times\MS{1}$ with no NS-NS 2-form fluxes.
Since $\lam\rightarrow\infty$ we view this as $k$ M5-branes in
M-theory on $\MR{2,1}\times \MT{3}\times\MS{1}\times \TBNT_q$.
Let $M_p$ be the 11-dimensional Planck scale.
The radius of $\MS{1}$ is, $R$. They are related according to,
$$
R = {\lam\over {m_s}},\qquad
M_p^3 = {{m_s^3}\over \lam}.
$$
The radius of $\TBNT_q$ is, $\rho$.

\Step{1}
Since, in the limit \gvul,
$$
M_p\rho = m_s \lam^{-1/3}\rho\rightarrow 0,
$$
we should view the fiber of the Taub-NUT
as the $11^{th}$ small dimension and convert to type-IIA
on $\MR{2,1}\times\MT{3}\times\MS{1}\times\MR{3}$.
We also have $k$ NS5-branes on $\MR{2,1}\times\MT{3}$
and $\TBNT_q$ became
$q$ D6-branes on $\MR{2,1}\times\MT{3}\times\MS{1}$.
The $\tw$-twists became RR 1-form Wilson lines along the
cycles of $\MT{3}$.
The string coupling constant is given by,
$$
\lam' = \lam^{-1/2}(m_s\TNS)^{3/2}\rightarrow 0.
$$
The new string scale is,
$$
M_s' = m_s^{3/2}\TNS^{1/2}\lam^{-1/2},
$$
and the radii of $\MT{3}$ satisfy,
$$
M_s' R_i = 
 m_s^{3/2}\TNS^{1/2}\lam^{-1/2} R_i\rightarrow 0.
$$
This means that we must perform T-duality on $\MT{3}$.

\Step{2}
After T-duality on $\MT{3}$ we obtain type-IIB on 
$\MR{2,1}\times\tilde{\MT{3}}\times\MS{1}\times\MR{3}$
with radii $\hat{R}_i$ which satisfy,
$$
M_s' \hat{R}_i = m_s^{-3/2}\TNS^{-1/2}\lam^{1/2} R_i^{-1}
       \rightarrow\infty.
$$
There are now
$k$ NS5-branes on $\MR{2,1}\times\tilde{\MT{3}}$
and $q$ D3-branes on $\MR{2,1}\times\MS{1}$.
At this point the $\tw$-twists became RR 2-form fluxes,
$$
\tw_i\epsilon_{ijk} = \int_{C_{jk}} B^{RR},\qquad i,j,k=1\dots 3,
$$
where  $C_{jk}$ is the 2-cycle made out of the $j^{th}$ and $k^{th}$
directions in $\MT{3}$.
The string coupling is now,
$$
\lam^{(2)} = {{\lam'}\over {m_s^{9/2}\TNS^{3/2}\lam^{-3/2} R_1 R_2 R_3}}
          = \lam m_s^{-3} R_1^{-1} R_2^{-1} R_3^{-1}
            \rightarrow\infty.
$$
This means that we must do S-duality.

\Step{3}
After S-duality we get type-IIB with $q$ D3-branes  and $k$ 
D5-branes in the same geometry.
The string coupling constant is now,
$$
\lam^{(3)} = \lam^{-1} m_s^3 R_1 R_2 R_3\rightarrow 0,
$$
and the string scale is,
$$
{M_s^{(3)}} 
= \lam^{-1}\TNS^{1/2}m_s^3 (R_1 R_2 R_3)^{1/2}.
$$
The radii satisfy,
$$
M_s^{(3)} \hat{R}_i 
 = \TNS^{-1/2}(R_1 R_2 R_3)^{1/2} R_i^{-1}
   \rightarrow 0,
$$
and the radius of $\MS{1}$ satisfies,
$$
M_s^{(3)} R = 
   m_s^2\TNS^{1/2}(R_1 R_2 R_3)^{1/2}\rightarrow \infty.
$$
At this point,
$$
\tw_i\epsilon_{ijk} = \int_{C_{jk}} B^{NSNS},\qquad i,j,k=1\dots 3.
$$
Since $M_s^{(3)}\hat{R}_i\rightarrow 0$, we must perform another
T-duality on $\MT{3}$. However, because of the NS-NS 2-form fluxes,
just as in \rCDS, another T-duality will not help.
Instead, let us do a T-duality on $\MS{1}$ which brings us to
the final setup of gauge theory on a non-commutative $\MT{4}$.

\Step{4}
After T-duality along $\MS{1}$ we get type-IIA with $k$ D6-branes
and $q$ D2-branes.
The string coupling is now,
$$
\lam^{(4)} = {{\lam^{(3)}}\over { M_s^{(3)} R }}
 =\lam^{-1}m_s\TNS^{-1/2}(R_1 R_2 R_3)^{1/2}\rightarrow 0,
$$
and $\hat{M}_s = M_s^{(3)}$.
The radii satisfy,
$$
\hat{M}_s\hat{R}_i =
  \TNS^{-1/2}(R_1 R_2 R_3)^{1/2} R_i^{-1}\rightarrow 0,
$$
and the radius of the $\MS{1}$ satisfies,
$$
\hat{M}_s \hat{R} = 
   m_s^{-2}\TNS^{-1/2}(R_1 R_2 R_3)^{-1/2}\rightarrow 0.
$$
At this point, the $\tw$-twists are still NS-NS 2-form fluxes.
We thus end up with a system of $k$ D6-branes on $\MT{4}\times\MR{2,1}$
and $q$ D2-branes which are points on $\MT{4}$. The radii of $\MT{4}$
are given, in terms of the 3 radii $R_i$ of the original
$\MT{3}$, as follows,
\eqn\hiluf{\eqalign{
\hat{R}_i &=\hat{M}_s^{-1}\TNS^{-1/2}
 (R_1 R_2 R_3)^{1/2}R_i^{-1},\qquad i=1,2,3,\cr
\hat{R}_4 &= \hat{M}_s^{-1}m_s^{-2}\TNS^{-1/2}(R_1 R_2 R_3)^{-1/2}.\cr
}}
Here $\hat{M}_s$ denotes the final type-IIA (with
the D2-branes and D6-branes) string scale.
The final string coupling constant is,
$$
\hat{\lam} = 
 \lam^{-1}m_s\TNS^{-1/2}(R_1 R_2 R_3)^{1/2}.
$$

Similarly, we can start with $S_A(k)$ with 3 $\etw$-twists.
By definition, this is $S_B(k)$ on the dual $\MT{3}$ with
3 $\tw$-twists.
We realize this in type-IIB on the background 
$\MR{2,1}\times\MT{3}\times \TBNT_q$ and $k$ NS5-branes
on $\MR{2,1}\times\MT{3}$.
As before, the fiber of the Taub-NUT space is denoted by $\TNS$.
We first perform S-duality to replace the NS5-branes with $k$ D5-branes.
At this point the $\etw$-twists are off-diagonal components of
the metric $g_{i9}$ with $i$ in the direction of $\MT{3}$ and
$9$ in the direction of the Taub-NUT fiber.
Then, we perform T-duality on the direction of $\TNS$ to obtain
type-IIA on $\MR{2,1}\times\MT{3}\times\MS{1}\times\MR{3}$
with $q$ NS5-branes
on $\MR{2,1}\times\MT{3}$ and $k$ D6-branes on 
$\MR{2,1}\times\MT{3}\times\MS{1}$.
The $\etw$-twists became NS-NS 2-form fluxes $B_{i4}$ where
$4$ is the direction of $\MS{1}$.
Then, we do T-duality on the
three directions of $\MT{3}$. We obtain $k$ D3-branes on
$\MR{2,1}\times\MS{1}$ and $q$ NS5-branes.
The $\etw$-twists are now off-diagonal components $g_{i4}$.
We then do another S-duality to get $k$ D3-branes and $q$
D5-branes and, finally, another T-duality on $\MT{3}$.
At this point we are back with $k$ D6-branes and $q$ D2-branes.
The $\etw$-fluxes are now NS-NS 2-form fluxes $B_{i4}$.

The moduli space is thus the same as the moduli space
of $q$ D2-branes  inside $k$ D6-branes on $\MT{4}$ with
NS-NS 2-form fluxes. In the case of $\tw$-twists,
these fluxes have both indices in the direction
of $\MT{3}\subset \MT{4}$. In the case of $\etw$-twists, the 
fluxes had one index in the direction of $\MT{3}$ and the other
index in the $4^{th}$ direction.
In the generic case, we have both $\tw$-twists and $\etw$-twists
simultaneously. The result is that the NS-NS 2-form flux
is nonzero for all 6 2-cycles of $\MT{4}$.
The string scale, string coupling, and
the parameters of the $\MT{4}$ are as calculated above.
We could in principle follow the chain of dualities above
with simultaneous $\tw$-twists and $\etw$-twists
but the intermediate steps would involve cumbersome 
non-linear expressions.

The moduli space of $q$ D2-branes inside $k$ D6-branes on $\MT{4}$
with NS-NS 2-form fluxes, and in the limit that the size of the
$\MT{4}$ vanishes,
was shown to be equivalent to the moduli space
of $k$ instantons of $U(q)$ gauge theory on a
non-commutative $\MT{4}$
\refs{\rDouglas-\rANS}.
It is likely that this result is true even for $\MT{4}$ of finite
size, because the size decouples by arguments as above.

In the next sections we will review the non-commutative geometry
and formulate a precise statement about the moduli space.

\newsec{Review of Noncommutative Gauge Theory}
In this section we will review the elements of non-commutative 
gauge theory which are relevant to our situation. 

Non-commutative gauge theory first entered string theory in 
\rCDS\  where it was shown to provide a matrix model for 
M-theory on a torus with the $C^{(3)}$ field turned on 
along the light-like circle. Subsequently, a lot of interesting 
work on this topic was done
\refs{\rDH-\rHVer}.
What we need here is not 
the connection to matrix theory but just the study of D-branes 
with a $B^{NS}$ fields turned on.

Consider type-IIA on $\MR{1,9-d} \times \MT{d}$ with $q$ D0-branes. 
The radii of $\MT{d}$ are called $R_i, i=1,..,d$, the string mass 
$m_s$ and the coupling $\lambda$. Furthermore let there be a 
constant $B^{NS}$ field along $\MT{d}$. Let 
\eqn\bfelter{
b_{ij} = \int_{ij} B^{NS},\qquad     i,j=1,\ldots,d
}
be the flux of $B^{NS}$ through the $\MT{2}$ spanned by 
directions $i,j$. The $b_{ij}$ are periodic with period 
$2 \pi$ due to the gauge invariance of $B^{NS}$. 

In \rCK\ this system was studied using the approach 
of \rWati. The result is that the 
low energy physics is described by a $d+1$ dimensional 
 $U(q)$ gauge theory on a dual torus, $\MHT{d} \times 
 \MR{0,1}$ with radii
\eqn\durad{
\tilde{R}_i = {1 \over m_s^2 R_i }
}
and gauge coupling 
\eqn\kob{
{1 \over g^2} = {m_s^{2d-3} R_1 \ldots R_d \over \lambda}.
}
The effect of $b_{ij}$ is to change the action. Every time 
two fields are being multiplied, the multiplication is with 
the $*$-product defined as,
\eqn\stjerne{\eqalign{
(\phi^{(2)} * \phi^{(1)})(x) &=
\left.
e^{-{{b_{ij}} \over {2 m_s^4 R_i R_j}}
(\partial^{(2)}_i \partial^{(1)}_j
-\partial^{(2)}_j \partial^{(1)}_i)}
    \phi^{(2)}(x_2)  \phi^{(1)}(x_1) \right|_{x^{(2)}=x^{(1)}=x},\cr
\partial^{(a)}_i &\equiv{\partial\over {\partial x^{(a)}_i}},\qquad a=1,2.
\cr
}}
The action is the usual gauge theory action just with this 
modification.

If there had been no $B^{NS}$-field the resulting $d+1$ dimensional 
gauge theory could have been obtained by performing 
T-duality along $\MT{d}$. The $q$ D0-branes would have turned into 
$q$ Dd-branes. The radii and gauge coupling of the $U(q)$ theory 
can be calculated in this way. The important point to remember 
is that the only change from having a $B^{NS}$-field is to 
change the product into eq.\stjerne.
The radii and gauge coupling are independent of $b_{ij}$. 
This result could not have been obtained by T-duality, 
since $B^{NS}$-fields change the formulas of T-duality and would 
have given other radii and gauge coupling.

There is another way of formulating this gauge theory. Instead of 
working with the $*$-product, eq.\stjerne, one can say that the 
torus $\MHT{d}$ is non-commutative. The algebra
 of functions on the torus is,$A$, is generated by $U_1, \ldots , 
U_d$ with relations 
\eqn\rela{
U_i U_j = U_j U_i e^{i b_{ij}}
}
The generalization of finite dimensional vector fields is 
finitely generated projective modules over $A$. Let $E$ be 
such a module. One can define connections,$\nabla$, and curvature 
$F_{ij}$ of this module \refs{\rCDS,\rANS}. One can define 
the Chern character of the module $E$
\eqn\chkar{
ch(E) = \sum_{k=0} {\hat{\tau} (F^k) \over (2\pi i )^k k!}
}
$\hat{\tau}$ is the trace on $End_A(E)$. $ch(E)$ 
can be regarded as an element in the cohomology, 
$H^{*}(\MT{d},{\bf C})$, of $\MT{d}$, the original 
torus. $ch(E)$ is not integral but there exists an 
integral cohomology class $\mu(E) \in H^{*}(\MT{d},{\bf C})$
 such that
\eqn\hele{
ch(E) = e^{{1 \over 2\pi}{\iota (b)}} \mu(E)
}
Here $\iota(b)$ denotes contraction with $b$ considered as an 
element of $H_{*}(\MT{d},{\bf C})$ \rANS. 

The mathematical fact that the module $E$ is 
characterized by integers is in 
exact agreement with our expectation from D-brane physics. 
Besides the $q$ D0-branes on $\MT{d}$ there could be any number 
of D2-branes, D4-branes , etc. wrapped on $\MT{d}$. These numbers 
are exactly given by $\mu(E)$.
$ch(E)$
measures the fact that D2-branes with $B^{NS}$-fields turned on 
have an effective D0-brane charge and the equivalent phenomena 
for other branes. Suppose for instance that only  $\mu_0$ and
$\mu_1$ are nonzero, then,
\eqn\eksempel{
ch_0 = \mu_0 + {b_{12} \over 2\pi} \mu_1,\qquad
ch_1 = \mu_1.
}
This equation reflects the fact that the number of D2-branes 
is unchanged by the presence of the $B^{NS}$-field but the 
number of D0-branes is shifted by the product of 
the number of D2-branes and the $B^{NS}$-field along the 
D2-branes.

\newsec{Noncommutative Instantons as the Moduli-space}
Let us now go back to our system of $q$ D2-branes inside 
$k$ D6-branes given above. They have a common $\MR{1,2}$. This 
is the space-time in which the 3 dimensional theory is living. 
The 3 dimensional theory has a low energy description as a sigma model 
with the moduli space of vacua as target space. 
The moduli space of vacua is a Hyper-k\"ahler manifold. The 
moduli space of vacua comes from the dynamics on the $\MT{4}$,  
which is the same as the dynamics of $q$ D0-branes in $k$ D4-branes on 
$\MT{4}$. The radii of the $\MT{4}$, 
$\hat{R}_1,\hat{R}_2,\hat{R}_3,\hat{R}_4$, and the string coupling
$\hat{\lam}$ and string scale $\hat{M}_s$
are given in terms of the parameters of 
the $S_A(k)$ compactification in \hiluf\ which we repeat here,
\eqn\otekhi{\eqalign{
\hat{R}_i &= m_s^{-3}\lam\TNS^{-1} R_i^{-1},\qquad i=1,2,3,\cr
\hat{R}_4 &= m_s^{-5}\lam\TNS^{-1}(R_1 R_2 R_3)^{-1},\cr
\hat{M}_s &= \lam^{-1}m_s^3\TNS^{1/2} (R_1 R_2 R_3)^{1/2},\cr
\hat{\lam} &=
  \lam^{-1}m_s\TNS^{-1/2}(R_1 R_2 R_3)^{1/2},\cr
}}
 Furthermore there is a $B^{NS}$-field turned on along $\MT{4}$,
\eqn\bfelter{\eqalign{
\int_{12} B^{NS} = \alpha_3, &\qquad
\int_{31} B^{NS} = \alpha_2, \cr
\int_{23} B^{NS} = \alpha_1, &\qquad
\int_{i4} B^{NS} = \eta_i,\qquad i=1,2,3.\cr
}}
but the vacuum structure of the vector-multiplets should be 
independent of $\TNS$ in this limit.

According to the above review of non-commutative geometry, 
the moduli space is equal to the moduli space of $k$ instantons 
in $U(q)$ gauge theory on a non-commutative torus, $\MHT{4}$,
with non-commutativity parameters equal to $\alpha_i$, $\eta_i$.
As explained above the radii and gauge coupling of this gauge theory 
are the same as if $\alpha_i = \eta_i =0$. Hence they can be found by 
T-duality on $\MT{4}$. By this T-duality one obtains $k$ D2-branes 
in $q$ D6-branes on $\MHT{4}$ of radii,
\eqn\radierto{
\tilde{R_1} = {\lam \over {m_s^3 R_2 R_3}},\,\,
\tilde{R_2} = {\lam \over {m_s^3 R_1 R_3}},\,\,
\tilde{R_3} = {\lam \over {m_s^3 R_1 R_2}},\,\,
\tilde{R_4} = {\lam \over m_s},
}
and string mass, $\tilde{m_s}$, and coupling, $\tilde{\lambda}$,
\eqn\vaerd{\eqalign{
\tilde{m_s} =
\hat{M}_s = \lam^{-1}m_s^3\TNS^{1/2} (R_1 R_2 R_3)^{1/2},\qquad
\tilde{\lambda} = \lam^{-1} m_s^3 \TNS^{3/2} (R_1 R_2 R_3)^{1/2}.
}}
In the $U(q)$ theory, this gives a gauge coupling of,
\eqn\qcdkob{
{1 \over g^2} = {\tilde{m_s}^3 \over \tilde{\lambda}} =
\lam^{-2} m_s^6 R_1 R_2 R_3.
}
Observe that $\TNS$ has dropped out of the radii and 
the gauge coupling.

 What about the limit $\lam\rightarrow\infty$ and $m_s$ fixed.
To see that the moduli space 
of vacua is well defined in this limit we should remember that 
scalar fields in three dimensions have dimension $\half$, if 
we want a standard kinetic term. We can either view the moduli space 
of vacua from the $U(q)$ gauge theory point of view or from the 
$U(k)$ theory on the D2-branes. {}From the last point of view 
the moduli space is the Higgs branch. The action of the $U(k)$ 
theory has a term,
\eqn\led{
\half {1 \over \tilde{\lambda}\tilde{m_s}} \int d^3x 
(\partial_\mu (\tilde{m_s}^2 X^i))^2
}
We define $\Phi^i = \tilde{\lam}^{-1/2}\tilde{m_s}^{3/2} X^i$. 
This $\Phi$ has a standard kinetic term,
\eqn\ledto{
\half \int d^3x (\partial_\mu \Phi^i)^2
}
The radii of the $\Phi^i$ are $R(\Phi^i) = 
\tilde{\lam}^{-1/2}\tilde{m_s}^{3/2}\tilde{R^i}$.
\eqn\endrad{\eqalign{
R(\Phi^1) = \sqrt{R_1 \over R_2 R_3},\qquad
R(\Phi^2) =& \sqrt{R_2 \over R_1 R_3},\qquad
R(\Phi^3) = \sqrt{R_3 \over R_1 R_2} \cr
R(\Phi^4) =& m_s^2 \sqrt{R_1 R_2 R_3}.
}}
We see that the limit $\lam \rightarrow \infty$ exists.
This last discussion was really superfluous. Since $S_A(k)$ only 
depends on the combination $m_s^2$ and does not feel $\TNS$, 
this had to be true. For finite $m_s\TNS$,
it could even be true for the full theory, not just the 
 moduli space of vacua. The effect of the twists is just to 
deform the moduli space and so does not change the fact that 
the moduli space is independent of $\TNS$ and has a limit when 
$\lam\rightarrow\infty$, keeping $m_s$ fixed.

We can also see from \endrad\ what happens in the limit of the $(2,0)$
theory. For this limit we take $m_s\rightarrow\infty$.
We find that the $\MT{4}$ degenerates to $\MT{3}\times \BR$.

Let us now be more precise about the space of instantons on 
a non-commutative $\MT{4}$. For this sake we will temporarily neglect the
 uncompactified directions and think of our system as $q$ D0-branes and 
k D4-branes on $\MT{4}$.
 According to the review of non-commutative geometry above, 
this is described by a gauge theory on the dual $\MHT{4}$ 
with non-commutativity parameters, $b_{ij}$, equal to the twists. 
 By gauge theory we really mean a projective module, $E$, which 
is characterized by 
\eqn\karbun{
\mu(E) = H^*(\MT{4},{\BZ}).
}
$\mu(E)$ has components in dimensions 0,2 and 4. $\mu_0 = q$ is 
the number of D0-branes on $\MT{4}$. $(\mu_1)_{ij}$ is the number of 
D2-branes in the $\MT{2}$ in direction $(i,j)$ with $i,j=1,2,3,4$. 
$\mu_2 = k$ is the number of D4-branes. So far we have not specified 
the number of D2-branes. Since we are interested in the low energy 
dynamics we should take the number of D2-branes to minimize the 
total energy in the D0,D2,D4 brane system. When $b_{ij}=0$ this 
is done by setting $\mu_1=0$, i.e. no D2-branes. Let us turn on 
$b_{12}$, say. {}From the formula
\eqn\heleto{
ch(E) = e^{{1 \over 2\pi}{\iota (b)}} \mu(E)
}
we get 
\eqn\mini{
(ch_1)_{34}= (\mu_1)_{34} + {b_{12} \over 2\pi} \mu_2
           = (\mu_1)_{34} + {b_{12} \over 2\pi} k.
}
To minimize the energy, $(ch_1)_{34}$ should be minimized. We see 
that when $b_{12} > {1 \over 2k} 2\pi$ we can lower the energy 
by taking $(\mu_1)_{34} = -1$. This phenomena divides the space 
of $b_{ij}$ into ``Brillouin'' zones. Each zone is a six dimensional 
cube of length $2\pi \over k$ in each direction. Inside a zone 
the low energy physics is described by the gauge theory corresponding 
to a module with the $\mu(E)$ which minimizes the energy. In crossing 
the boundary between 2 zones, $\mu(E)$ jumps. 

We also see another interesting phenomena. Whenever ${b_{12} \over 2\pi}k$ 
 is an integer we have $(\mu_1)_{34} =- {b_{12} \over 2\pi}$ and 
hence $(ch_1)_{34} =0$. This means that $ch(E)$ is nonzero only in 
dimensions 0 and 4 (We are keeping all other components of $b_{ij}=0$. 
Only $b_{12}= n {2\pi \over k}$). This is exactly like the pure D0,D4 
system with no $B^{NS}$-field. This system has a phase where the 
D0-branes and D4-branes are separated. To reach this phase the 
system has to go through zero-size instantons. We thus conclude that 
whenever $b_{12}= n {2\pi \over k}$,$n \in {\BZ}$ there is another phase. 
Of course, there is nothing special about $b_{12}$. Similar statements 
could be made for the other 5 components of $b_{ij}$ and even for all 
of them simultaneously. The point is that for each center of 
the ``Brillouin'' zone there is another branch emanating from a 
locus on the Coulomb branch. It emanates from the points on the  
Coulomb branch where some instantons have shrunk to zero size. 
The other phase consists of the $k$ D4-branes with $-n$ D2-branes 
inside moving away from the $q$ D0-branes. Let us calculate the  
dimension of this branch. Suppose first $n=1$, so there are $k$
D4-branes with $-1$ D2-brane inside (equivalently 1 anti D2-brane). 
This system has a bound state. It is not marginally bound. The system 
has an 8 dimensional moduli space. To see this we should really 
remember that it is really $k$ D6-branes with $-1$ D4-brane. 
4 of the dimensions are $U(1)$ Wilson lines on the $\MT{4}$. 
They are center of mass coordinates and are always present. We are 
not interested in these. The other 4 are 3 transverse positions and 
the dual photon in 3 dimensions. We conclude that the other phase is 
4 dimensional. Furthermore it emanates from a point 
on the Coulomb branch, since all instantons have to shrink on top 
of each other. The only freedom is the point where they shrink, but 
that is a center of mass degree of freedom which we ignore.

Let us now take $n$ to be generic. Let $g= gcd(n,k)$. The system of 
$n$ D2-branes inside $k$ D4-branes can split into $g$ separate systems. 
The dimension is thus $8g-4$, subtracting the center of mass again. 
It emanates from the Coulomb branch on a locus of dimension $4g-4$.

The special case of $q=1,k=2$ was studied in detail in 
\rCGK. Here it was found that there was another phase of 
dimension 4 for $\alpha = \pi$. We see that this agrees exactly 
with what was found here. However we get a much clearer 
picture of the other branch. In the next section we will 
understand these branches from a field theory point of view.

\subsec{Phase Transitions from the Gauge Theory}
With generic twists (non-commutativity parameters), the moduli-space
that we obtain is smooth.
However, for special values of the twists the moduli space has ADE-type
singularities. We would now like to explain the origin
of some of these singularities.

$S_B(k)$ is a gauge theory at low energies. Let us study it with an 
$\alpha$-twist along one circle and no twist along the other 2 circles. 
 Since there is a circle without twist we can T-dualize
on that direction to $S_A(k)$, so 
these remarks apply to $S_A(k)$ as well. We want to reproduce the 
existence of other branches of the moduli space. 
 For a related discussion see \rNek.

The fields in 6 dimensions are a $U(k)$ vector-multiplet and an 
adjoint hypermultiplet. In 3 dimensions there is a tower of $U(k)$ 
vector-multiplets with masses $({n_1 \over R_1},
{n_2 \over R_2},{n_3 \over R_3})$, $n_i \in {\BZ}$ 
and a tower of adjoint hypermultiplets with masses
$({n_1 - {\alpha \over 2\pi} \over R_1},
{n_2 \over R_2},{n_3 \over R_3})$, $n_i \in {\BZ}$.
We remember that a mass in $\SUSY{4}$ theories in
 3 dimensions is specified by 3 numbers.
The moduli space is $4k$-dimensional including the center 
of mass degrees of freedom. On the Coulomb branch the $U(k)$ is 
broken to $U(1)^k$. Each adjoint hypermultiplet splits into 
$k^2$ hypermultiplets of the following charges. There are 
hypermultiplets with charge $(0,\ldots,0)$, and
there are $k$ hypermultiplets  with 
charges $(1,-1,\ldots,0)$ plus permutations. There 
is a total of $k(k-1)$ of these. 
Some of these hypermultiplets can become massless on the 
Coulomb branch. For that to happen we have to turn on a 
Wilson line, $A_1$, along the first circle and set the other
$3k$ moduli zero. $A_1$ has the form 
\eqn\matriks{
A_1 = \left(\matrix{ a_1 & 0 & \ldots & 0 &0 \cr
                       0 & a_2 & \ldots & 0 &0 \cr
                       \vdots & \vdots&\ddots& \vdots&\vdots \cr
                     0   & 0 & \ldots & a_{k-1} & 0 \cr 
                     0   & 0 & \ldots & 0 & a_{k} \cr} \right)
}
The tower of hypermultiplets is now as follows.
There are $k$ of charge $(0,\ldots,0)$ with mass 
$({n_1 - {\alpha \over 2\pi} \over R_1},
{n_2 \over R_2},{n_3 \over R_3})$ and for every $i\ne j$ there is
a hypermultiplet with charge
$(0,\ldots,1,\ldots,-1,\ldots,0)$ plus permutations
with the 1 on the $i^{th}$ place and the -1 on the $j^{th}$ place. 
It has a  mass $({n_1 - {\alpha \over 2\pi} + {a_i \over 2\pi} 
- {a_j \over 2\pi} \over R_1},
{n_2 \over R_2},{n_3 \over R_3})$.
The uncharged ones never become massless, as long as the twist is 
not a multiple of $2\pi$. The charged ones become massless if 
\eqn\masseloes{
n_1 - {\alpha \over 2\pi} + {a_i \over 2\pi} 
- {a_j \over 2\pi} = n_2 = n_3 =0.
}
Now it is easy to make some of them massless by choosing $A_1$ 
appropriately. However to have a Higgs branch we need to have 
non trivial solutions to the D-flatness equations. For 
hypermultiplets charged under a $U(1)^r$ group there should 
be at least $r+1$ of them to have a non trivial solution. We thus 
need to find a number of massless hypermultiplets which is bigger than 
the number of $U(1)$'s under which they are charged. No hypermultiplets 
are charged under the diagonal $U(1)$. Let us first find a situation 
of k massless hypermultiplets which are charged under $U(1)^{k-1}$.
The hypermultiplet of charge $(1,-1,0,\ldots,0)$ is massless 
if,
\eqn\maslet{
n_1=0,\qquad a_1-a_2= \alpha.
}
The one of charge $(0,1,-1,0,\ldots ,0)$ is massless if,
\eqn\maslto{
n_1=0,\qquad a_2-a_3= \alpha
}
and so on, up to the multiplet of charge $(0,\ldots,0,1,-1)$ 
which is massless if,
\eqn\masltre{
n_1=0,\qquad a_{k-1}-a_k= \alpha
}
This gives $k-1$ massless hypermultiplets. To have one more we need 
$(-1,0,\ldots,0,1)$ to be massless. This is the case if,
\eqn\maslfire{
{\alpha \over 2\pi} = {a_k-a_1 \over 2\pi} + n_1,
}
for some integer $n_1$. Now,
\eqn\maslfem{
a_k - a_1 = (a_k -a_{k-1}) + \ldots + (a_2 -a_1) = -(k-1)\alpha
}
so we need ${{k\alpha} \over {2\pi}}$ to be an integer. So for $\alpha 
= {2\pi \over k}$ we have another phase of dimension 4. The 
dimension is 4 because there are $k$ massless hypermultiplets
 each having 4 scalar fields and the D-flatness conditions 
remove $4(k-1)$ dimensions leaving 4 real dimensions. This phase 
 agree agrees exactly with the exact result from the previous section. 
We thus see that a naive field theory treatment, keeping all 
Kaluza-Klein modes, reproduces the result. 
This phase emanates 
from the Coulomb branch whenever $a_i = a_{i-1} = \alpha$ 
as we saw above. This fixes the $a_i$ up to an overall shift. 
The overall shift is the $U(1)$ part which we discard anyway. 
This shows that the other phase emanates from one
particular point on the Coulomb branch.
Note that the field theory treatment is justified when $M_s R_i\gg 1$.

More generally, let us take $\alpha = n {2\pi \over k}$ and 
$g = gcd(n,k)$. Now we can play the same game as above but within 
$g$ blocks of the $U(k)$ matrix of size $k \over g$. We thus get 
g sets of $k \over g$ massless fields. Each set is charged under 
a $U(1)^{{k \over g} -1}$ subgroup.
This gives a $4g$ dimensional phase 
emanating from a locus on the Coulomb branch. This locus has 
dimension $4g-4$. The $4g$ comes from the diagonal $U(1)$ in each 
of the $g$ blocks. The center of mass is subtracted again. 
This branch has a total dimension of $4g+4g-4 = 8g-4$. We again 
find agreement with the exact result described previously. 

The branches described above are the only ones coming from the 
naive field theory description besides the cases $\alpha = 2 \pi n$, 
$n \in {\BZ}$ which behave like $\alpha =0$.


\newsec{The 3+1D limit}
In this section we will explain how to obtain the 3+1D Seiberg-Witten
curves of the various theories
compactified on $\MT{2}$ with a twist. This time we only
have two independent
$\tw$-twists corresponding to the two cycles of $\MT{2}$.
The way to obtain the 3+1D SW curves is to start with the moduli
space of the theory compactified on $\MT{2}\times\MS{1}$ where
$\MS{1}$ is of radius $R$ and take the limit $R\rightarrow\infty$.
Let the 2+1D
hyper-K\"ahler moduli space be of dimension $4n$.
In the limit $R\rightarrow\infty$, it 
can be written as a fibration of $T^{2n}$ over a base of dimension $2n$.
In the decompactification limit the fiber $T^{2n}$ shrinks to zero.
We interpret it as the Jacobian variety of a Riemann surface of genus
$n$ which varies over the base. This will then be the Seiberg-Witten curve
(see \rSWGDC).
Starting with the Blum-Intriligator little-string theories
of $k$ NS5-branes at an $A_{q-1}$ singularity compactified on $\MT{2}$ 
with twists we can get, in appropriate limits, a 3+1D gauge theory
with,
$$
SU(k)_1\times\cdots\times SU(k)_q,
$$
and massive adjoint hyper-multiplets in consecutive $(k,\bar{k})$
representations.
The Seiberg-Witten curves for these models have been derived in \rWitFBR.
As we will show below, we can reproduce these curves by taking the
appropriate  decompactification limit of 
the moduli space of $k$ $U(q)$ instantons on the non-commutative $\MT{4}$.

To start, we will recall how the reduction of the untwisted compactified
Blum-Intriligator theories works.

\subsec{From instantons to quiver gauge theories}
When we set all the $\tw$-twists to zero we obtain the statement
that the Coulomb-branch moduli space of the theories of
$k$ NS5-branes on an $A_{q-1}$ singularity, compactified on 
$\MT{3}$ is the same as the moduli space of $k$ ordinary instantons
with a $U(q)$ gauge group on $\MT{4}$.
This result has already been established in \refs{\rIntNEW, \rGanSet}.
Suppose we compactified on $\MT{3} = \MT{2}\times \MS{1}$ and take
the radius of $\MS{1}$, $R\rightarrow\infty$.
It can be checked (see \endrad)
that the auxiliary $\MT{4}$ becomes a product
$\MT{2}_B\times \MT{2}_F$.
The complex structure of $\MT{2}_F$ and $\MT{2}_B$
are fixed as $R\rightarrow\infty$ while the area of $\MT{2}_B$ is
proportional to $R$ and the area of $\MT{2}_F$ is proportional to $R^{-1}$.
Now take a particular gauge configuration corresponding to an instanton
of $U(q)$ with instanton number $k$.
We can encode the information in the instanton as follows
(see \refs{\rFMW,\rBJPS}).
At a local point on the base, the gauge field reduces to two commuting 
$U(q)$ Wilson lines on the fiber. We can describe them uniquely
as $q$ points on the dual $\MHT{2}$ of the fiber.
These $q$ points vary over the base $B$. The instanton equations
imply that they span a holomorphic curve $\Sigma_g$ of genus 
$g = q k + 1$. $\Sigma_g$ is called the {\it ``spectral curve''}.
 To completely describe the instanton we also need to
describe a line bundle over $\Sigma_g$ which corresponds to a point
in the Jacobian of $\Sigma_g$ (recall that the Jacobian of a genus
$g$ curve is $T^g$). The line bundle is called the
{\it ``spectral-bundle''}.
Alternatively, we can represent the moduli space of $U(q)$ instantons
at instanton number $k$ on $B\times F$ as the moduli space of 
$q$ D6-branes wrapped on $B\times F$ with $k$ D2-branes.
The curves are obtained by T-duality along the two directions 
of $F$. We obtain a D4-brane wrapped on a curve $\Sigma_g$
of homology cycle $q\lbrack B\rbrack + k \lbrack F\rbrack$.
 The curve $\Sigma_g$ is the Seiberg-Witten curve
of the point in the moduli space. It intersects a generic fiber $F$
in $q$ points and a zero section of the base $B$ at $k$ points.
It is also easy to see that as the base $B$ decompactifies to
$\MS{1}\times \MR{1}$ we reproduce exactly the curves from the
brane construction of \rWitFBR\ for the quiver gauge theory.

\subsec{The r\^ole of the non-commutativity}
Now let us repeat the same procedure but with two non-commutativity
parameters $\tw_1$ and $\tw_2$. We can take $\tw_1$
to be along the first cycle of the base $B=\MT{2}$ and the first
cycle of the fiber $F=\MT{2}$ and we take $\tw_2$ to be along the
second cycle of the base $B$ and the first cycle of the fiber $F$.
The $\eta$-twists will similarly correspond to non-commutativity
along the second cycle of $B$ and one of the two cycles of $F$.

To translate this to the curve $\Sigma_g$ we take the system of
$q$ D6-branes and $k$ D2-branes and put in NSNS 2-form fluxes
according to the non-commutativity parameters. After T-duality
along $F$ The NSNS fluxes become components of the metric $G_{IJ}$.

As a result, we obtain a tilted $\MT{4}\equiv\MR{4}/\Lambda$,
where $\Lambda$ is a lattice spanned by the following vectors:
\eqn\ves{\eqalign{
\ve_1 &= (1,0,0,0),\cr
\ve_2 &= (\tau_1,\tau_2,0,0),\cr
\ve_3 &= (\tw_1 + \etw_1\tau_1,\etw_1\tau_2,\chi,0),\cr
\ve_4 &= (\tw_2 + \etw_2\tau_2,\etw_2\tau_2,\chi\rho_1,\chi\rho_2).\cr
}}
Here, $\tau\equiv \tau_1 + i\tau_2$ is the complex structure
of $\MT{2}_F$, $\rho\equiv \rho_1 + i\rho_2$ is the complex 
structure of $\MT{2}_B$,
and,
$$
\chi = m_s (\tau_2\rho_2)^{-1},
$$
so that the overall volume of the unit  cell will be $m_s^2$.
We will denote the coordinates in $\MR{4}$ by $(x_1,x_2,x_3,x_4)$.
The D2 and D6 branes became a single D4-brane in the homology class,
$$
\lbrack \Sigma\rbrack = 
q\lbrack B'\rbrack + k \lbrack F'\rbrack.
$$
Here,
\eqn\fpbp{\eqalign{
 F' &\equiv \{s \ve_1 + t\ve_2\, |\, 0\le s,t\le 2\pi\},\cr
B' &\equiv \{s \ve_3 + t\ve_4\, |\, 0\le s,t\le 2\pi\}.\cr
}}
are two faces of $\MT{4}$.
Similarly to \rWitFBR\ the D4-brane will find a minimal-area surface in
this homology class.
In the complex structure given by,
$$
z = x_1 + i x_2,\qquad w = x_3 + i x_4,
$$
the cohomology class $\omega\in H^2(\BZ)$ which is Poincar\`e dual
to $\lbrack\Sigma\rbrack$ will, generically, be a mixture of
$(1,1)$, $(0,2)$ and $(2,0)$ forms. However, it is always possible to
find a complex structure (with respect to the flat metric) for which
$\omega$ is entirely a $(1,1)$ form. In this complex structure
the $\MT{4}$ is ``algebraic'' (see p315 of \rGriHar).
Given the complex structure, it is possible to write down
the curve $\Sigma$ as the zero locus of a $\theta$-function on
$\MT{4}$. These $\theta$-functions are the sections
of the line-bundle corresponding to $\lbrack\Sigma\rbrack$
and depend on $k q$ parameters which are the moduli (see \rGriHar\
for further details).

It is easy to see that the ``elliptic-models'' of \rWitFBR\
are recovered in the special limit in which we get a gauge
theory with massive hyper-multiplets.
In this case $\tau\rightarrow\infty$ and
there are no $\eta$-twists. The fiber $F'$ is replaced
with a strip $\MS{1}\times\MR{1}$. The class $\lbrack\Sigma\rbrack$
is analytic (i.e. the class $\omega$ is a $(1,1)$ 2-form)
and the Seiberg-Witten curves of \rWitFBR\ are recovered.


\newsec{Another Look at the $\etw$-twists}

In this section, we write explicitly the solution for type-IIA
(or type-IIB) theory, with both $\tw$-twists and $\etw$-twists
turned on. 
These solutions should be interpreted as string world-sheet
$\sigma$-models with a $B$-field.

We will start with a Taub-NUT space without NS5-branes.
It is straightforward to define  the $\tw$-twist.
One starts with some given background, which
is a principal $U(1)$ bundle cross a torus $\MT{d}$. 
Locally, the $\tw$-twist is just the change of coordinate in the
$\MS{1}$ fiber of the Taub-NUT space,
 of the form $y\to y+\sum\tw_I\psi^I$.
$y$ is the coordinate on the circle (see \tnmet) and $\psi^I$
is the coordinate on $\MT{3}$ ($I=1,2,3$).
Since it is
just the change of variables, the string theory equations of
motion are trivially satisfied. But globally, this is not
a valid coordinate transformation, since $\tw_I\psi^I$ is
not a periodic function on $\MT{3}$ modulo $2\pi$. Therefore,
we get a different background --
we call it the $\tw$-twisted background.
As for $\etw$-twists, they are related to $\tw$-twists by 
T duality in $\MT{3}$.

We will construct the background with both $\tw$ and $\etw$ twists
turned on in the following way. We first consider the background
containing Taub-NUT space cross a three-torus, without any twists.
We introduce $\tw$-twists along the three-torus, with the parameters
$\etw_I$. Then, we make a T-duality transformation, and get a background
with $\etw$-twists. This new background is again a $U(1)$ bundle
cross a (dual) torus, and we now $\tw$-twist it. In this way,
we get a background with both $\tw$-twists and
$\etw$-twists.

Let us do it explicitly.
Start with $\MR{1,2}\times \TBNT(\rho)\times \MT{3}$. The metric is:
\eqn\metriczero{\eqalign{
ds^2=&\rho^2 U_{[\rho]}(|\vec{r}|) {\cal A}^2
     + U_{[\rho]}(|\vec{r}|)^{-1} (d\vec{r})^2 
     \cr &
+ g_{IJ} d\psi^I d\psi^J - dx_0^2+dx_1^2+dx_2^2,\cr
}}
where we have denoted 
\eqn\defUr{
\Ur{\rho}\equiv\left( 1+{\rho\over 2 |\vec{r}|}\right)^{-1}
}
and $\cal A$ is the connection one-form
${\cal A}=dy-\vec{A}\cdot d\vec{r}$.
Also, we turn on the following $B$ field:
\eqn\Btostart{
B=b_{IJ} d\psi^I\wedge d\psi^J
}
We wish to introduce $\tw$-twists with the parameter $\etw_I$.
As was explained above, this means just the change of variables
$y\to y-\etw_Id\psi^I$. This amounts to replacing ${\cal A}^2$
with $({\cal A}-\etw_Id\psi^I)^2$ in \metriczero.

Now we make three T-dualities. We do this by the standard 
technique of treating $V^I_\a\equiv \px{\a}\psi^I$ (where
$\a$ is a string world-sheet coordinate) as an independent
variable and inserting a Lagrange multiplier, $\tilde{\psi}_I$, for,
$\px{\lbrack\a}V_{\b\rbrack}^I$.
We get the following metric:
\eqn\metricone{\eqalign{
ds^2=&{\rho^2 \Ur{\rho} \over 1+(\etw,\etw) \rho^2 \Ur{\rho}}
         ({\cal A}-b^{IJ}\etw_I d\tilde{\psi}_J)^2 
      + \Ur{\rho}^{-1}(d\vec{r})^2
      \cr
&+ l_s^4\left(g^{IJ}- 
       {\rho^2 \Ur{\rho}\over 1+\rho^2 (\etw,\etw)\Ur{\rho}}
      \etw^I \etw^J \right) d\tilde{\psi}_I d\tilde{\psi}_J            
      -dx_0^2 +dx_1^2 +dx_2^2,\cr
}}
with the notation,
$\etw^I=g^{IJ}\etw_J$, $(\etw,\etw)=\etw_I\etw^I$,
and $g^{IJ}+b^{IJ}$ is the matrix inverse to
$g_{IJ}+b_{IJ}$.
Also, we have the following $B$ field:
\eqn\bfield{
\matrix{
B=-{\rho^2 \Ur{\rho}\over 1+\rho^2 (\etw,\etw)\Ur{\rho}}
  \etw^I d\tilde{\psi}_I \wedge ({\cal A}-b^{JK}\etw_J d\tilde{\psi}_K)+
   b^{IJ}d\tp_I\wedge d\tp_J 
}}
Notice that 
\eqn\secondrho{
{\rho^2 \Ur{\rho} \over 1+(\etw,\etw) \rho^2 \Ur{\rho}}=
{\rho^2\over 1+(\etw,\etw)\rho^2} \Ur{{\rho\over 1+(\etw,\etw)\rho^2}}
}

If we start with a non-degenerate torus and a very small coupling constant,
then T-duality gives us back a very small coupling constant.

Now we $\tw$-twist this background. Again, $\tw$-twisting is just
a replacement,
$$
{\cal A}\to {\cal A}-\tw^Id\tp_I,
$$
in all the formulas for the metric and the $B$ field.
It is convenient
to absorb $b^{IJ}\etw_I d\tp_J$ into $\tw^I d\tp_I$. 
Then, the background fields are:

\eqn\both{
\matrix{
ds^2&=&
R^2(|\vec{r}|)({\cal A}-\tw^I d\tp_I)^2+
\Ur{\rho}(d\vec{r})^2+(dx^{\mu})^2+
l_s^4 G^{IJ}(|\vec{r}|)d\tp_I d\tp_J, \hfill
\cr
&&
\cr
B&=&
({\cal A}-\tw^I d\tp_I)\wedge B^J d\tp_J + B^{IJ}d\tp_I\wedge d\tp_J
\hfill
}}
where
\eqn\defs{
\matrix{
R^2(|\vec{r}|)&=&{\rho^2\Ur{\rho}\over 1+(\etw,\etw)\rho^2\Ur{\rho}} 
\hfill\cr
G^{IJ}(|\vec{r}|)&=&g^{IJ}-
{\rho^2\Ur{\rho}\over 1+(\etw,\etw)\rho^2\Ur{\rho}}\etw^I\etw^J 
\hfill\cr
B^I(|\vec{r}|)
&=&{\rho^2\Ur{\rho}\over 1+(\etw,\etw)\rho^2\Ur{\rho}}g^{IJ}\etw_J
\hfill\cr
B^{IJ}&=&b^{IJ}\hfill
}}
Also, the dilaton is not constant.
Let $\lambda$ be the string coupling at $|\vec{r}|\to\infty$.
Then, the string coupling at finite $|\vec{r}|$ is:
\eqn\StringCoupling{
\lambda(|\vec{r}|)=\lambda\sqrt{1+\etw^2\rho^2\over 
1+\etw^2\rho^2\Ur{\rho}}
}


The metric \both\ is not, strictly speaking,  Hyper-K\"ahler.
Indeed, although it does have three complex structures, they are
not covariantly constant with respect to the standard 
covariant derivative. But they must be covariantly constant, if we modify
$\Gamma_{\mu\nu}^{\rho}$ with the torsion, proportional to $H=dB$.

We want to study the moduli space of the theory on the NS5-brane,
sitting at $\vec{r}=0$ in this background.
As we remarked in section (2), the NS5-brane has a size of $l_s$
and, although it is very heavy, it could affect the metric.
We will explore  this later in this section. For now, we will
assume that it is safe to forget about the NS5-brane.
To study the moduli space, we perform
the chain of dualities. It is most convenient to think of these
dualities as acting on the asymptotic
 ($\vr\to\infty$) values of the fields.
Therefore, we would like to discuss how the background fields near
the position of the NS5-brane ($\vr\to 0$) are related to the asymptotic
values of the fields at $\vr\to\infty$.

Let us look first at the geometry near the origin in $\MR{3}$.
 {}From \metricone\ and \secondrho\ we see that the geometry becomes flat
when the following two conditions are satisfied:
\eqn\twocond{
|\vec{r}|\ll\rho \;\;\;\; {\rm and} \;\;\;\;  
|\vec{r}|\ll{\rho\over 1+(\etw,\etw)\rho^2}
}
In this limit, we have just $\MR{1,6}\times \MT{3}$
with the metric
\eqn\metricloc{
ds^2=(dx^{\mu})^2 + 
|d(e^{i\tw^J\tp_J}z_1)|^2+
|d(e^{-i\tw^J\tp_J}z_2)|^2 + g^{IJ} d\tp_I d\tp_J
}
The $B$ field becomes:
\eqn\bfieldloc{
B= -\etw^I d\tp_I \wedge {\rm Im} (z_1^* dz_1+ z_2^* dz_2)+
b^{IJ}d\tp_I\wedge d\tp_J 
}
We wish to study the moduli space for the NS five-brane 
sitting at $\vec{r}=0$. Notice that the transversal 
fluctuations of this five-brane at energy scale $\simeq m_s^2$
have the characteristic size $\Delta X^{\perp}\simeq \lambda l_s$.
If we take $\rho\simeq l_s$ and general $\etw$, then both of the
inequalities \twocond\ are satisfied for 
$|\vec{r}|\equiv \Delta X^{\perp}$. This suggests that the parameter
$\rho\simeq l_s$ actually does not affect the moduli space.
The reason why it might be not true is that the transversal
size of the NS5-brane is, actually, of the order $l_s$. Therefore
the curvature of the background should, presumably, affect the physics
even in the limit $\lambda\to 0$. 
The answer we will get  shows that the moduli space
does not really depend on $\rho$.

Now let us look at the fields at infinity. They are given by the
formulae \both\ and \defs\ with $\vr=\infty$. We will denote
the limits of $R^2(\vr)$, $G^{IJ}(\vr)$
and $B^I(\vr)$ as $\vr\to\infty$
by $R^2$, $G^{IJ}$ and $B^I$. It is convenient to have
a dictionary relating the fields at $\vr=\infty$ with the fields
at $\vr=0$. Let us first summarize our notations.
We have already introduced the matrices $g_{IJ}$, $b_{IJ}$,
$g^{IJ}$ and $b^{IJ}$ satisfying:
$$ (g^{IJ}+m_s^2 b^{IJ})(g_{JK}+l_s^2 b_{JK})=\delta^I_K $$
We have also introduced $G^{IJ}$ and $B^{IJ}$ in \both.
Now, we define $G_{IJ}$, $B_{IJ}$, $g^{-1}_{IJ}$ and
$G^{-1}_{IJ}$ in the following way:
\eqn\newdefs{
(G_{IJ}+B_{IJ})(G^{JK}+B^{JK})=\delta_I^K,\;\;\;
g^{-1}_{IJ}g^{JK}=\delta_I^K,\;\;\;
G^{-1}_{IJ}G^{JK}=\delta_I^K
}
Then, we have the following dictionary, relating asymptotic
background to the local background:
\eqn\dictionary{\eqalign{
\rho^2=R^2+(B,B), &\qquad R^{-2}=\rho^{-2}+(\etw,\etw),\cr
g^{IJ}=G^{IJ}+R^{-2}B^IB^J, &\qquad
         G^{-1}_{IJ}=g^{-1}_{IJ}+\rho^2\etw_I\etw_J,\cr
\etw_I={R^{-2}G^{-1}_{IJ}B^J\over 1+R^{-2}(B,B)}, &\qquad
         B^I={\rho^2\over 1+\rho^2(\etw,\etw)} g^{IJ}\etw_J,\cr
B^{IJ} &=b^{IJ}.\cr
}}
The local value, $\lambda_0$,
of the string coupling is related to the asymptotic
value $\lambda$ by the formula which follows from \StringCoupling:
\eqn\Coupling{
\lambda_0^2=(1+(\etw,\etw)\rho^2)\lambda^2
}


\subsec{The chain of dualities.}

We start by replacing the Taub-NUT circle with the M-theory
circle. We get a D6-brane wrapped on $\MT{4}$, with the NS5-brane
on top of it. 

At this point it is useful that we remember how the fields
of type-IIA theory are related to the fields of M-theory.
M-theory on a $U(1)$ bundle is type-IIA on the base of this bundle.
Suppose that the action of $U(1)$ is associated to the vector field
$v$. 
The M-theory three-form $C_M$ splits as follows:
\eqn\splitC{
C_M=\pi^* A^{(3)}+ {\cal A}\wedge \pi^* B
}
Also, we choose some local trivialization, and define the connection
one-form $A^{(1)}$ on the base, $dA^{(1)}={\cal F}$
(${\cal F}$ is the curvature two-form on the base,
$d{\cal A}=\pi^*{\cal F}$).  It should be identified
with the RR one-form  $C^{(1)}$ of type-IIA. Also, $B$ should be identified 
with the $B$ field of type-IIA (this follows from its coupling to the
fundamental string). What is the relation between $A^{(3)}$ and the
Ramond-Ramond three-form $C^{(3)}$ of type-IIA? 
Let us remember the general formula for the couplings of the Ramond-Ramond 
fields to the D-brane \rDouglas:
\eqn\couplings{
S_{RR}=\int\mu_p C\wedge {\rm tr} e^{F-B}
}
 For example, for the D2 brane we get:
\eqn\Dthree{
S_{RR}=\mu_2 \int C^{(3)}-C^{(1)}\wedge (B-F)
}
Here $C^{(1)}$ should be identified with the connection
one-form, ${\cal A}=d\phi+C^{(1)}$.
We have to keep in mind that various forms participating in this formula
are, in general, subject to gauge transformations.
 For example, under the gauge transformation $C^{(1)}\to C^{(1)}-d\psi$
we should have $C^{(3)}\to C^{(3)}-d\psi\wedge B$ (this is needed for
the coupling \Dthree\ to be correctly defined). This suggests that
\eqn\whatisC{
C^{(3)}=A^{(3)}+C^{(1)}\wedge B
}
(that is, $C_M=\pi^* C^{(3)}+d\phi\wedge \pi^* B$.) 
We may derive how Ramond-Ramond fields transform under T duality 
from their coupling to D branes. It follows that 
$Ce^{-B}$ transforms as a spinor of $O(d,d,{\bf Z})$.
Notice that
\eqn\CemB{
Ce^{-B}=A^{(1)}+A^{(3)}+{\rm forms\;\; of \;\; higher \;\; rank}.
}

Let us return to our dualities.
We assume that the M Theory circle 
in our original configuration has radius $S=\lambda l_s$,
where $l_s$ is the string scale in the configuration we start with,
and $\lambda$ is the original coupling constant (which has to be very
small, if we want to get Little String Theory on NS5 brane).
The three-form of M Theory is read from \both:
\eqn\Mthreeform{
C_M=({\cal A}-\tw^Id\tp_I) \wedge
B^Jd\tp_J\wedge d\theta+  B^{IJ}d\tp_I\wedge d\tp_J \wedge d\theta
}
If we now treat  the Taub-NUT circle as the M-theory circle, we
get \splitC\ with 
$$
A^{(3)}= B^{IJ}d\tp_I\wedge d\tp_J \wedge d\theta,\qquad
B=B^Id\tp_I\wedge d\theta.
$$
(Notice that ${\cal A}-\tw^I d\tp_I$ is just the connection 1-form after 
$\tw$-twist.)

In the new type-IIA theory, obtained by compactifying M Theory
on the Taub-NUT circle, we have the following asymptotic
values of the background fields:
\eqn\bgnd{\eqalign{
ds^2&=S^2 d\theta^2 + l_s^4 G^{IJ} d\tp_I d\tp_J
       +d\vec{r}^2 +(dx^{\mu})^2,\cr
B&=B^I d\tp_I\wedge d\theta,\cr
Ce^{-B}&=\tw^I d\tp_I+
d\theta\wedge B^{IJ}d\tp_I\wedge d\tp_J.\cr
}}
(We have used \CemB\ to find $Ce^{-B}$ in type-IIA.)
The new string length is:
\eqn\newls{
l_1^2={S\over R}l_s^2 = \lambda_0 {l_s^3\over\rho}
}
and the new string coupling constant is:
\eqn\couplingone{
\lambda_1=\left({R\over l_s}\right)^{3/2}{1\over\sqrt{\lambda}}
}
Making three $T$ duality transformations along $\MT{3}$, we get:
\eqn\bgndone{\eqalign{
ds^2&={l_s^2\lambda_0^2\over\rho^2}\left[\rho^2 d\theta^2 +   
      G^{-1}_{IJ} d\psi^I d\psi^J 
     +2 G^{-1}_{IJ} B^I d\psi^J d\theta \right]
     +d\vec{r}^2 + (dx^{\mu})^2,\cr
B^{RR}&=\tw^I\epsilon_{IJK}d\psi^J\wedge d\psi^K
+d\theta\wedge\epsilon_{IJK}B^{IJ}d\psi^K,
\cr
B^{NS}&=0,\cr
}}
with the string coupling constant,
\eqn\couplingtwo{
\lambda_2={\lambda\over l_s^3\sqrt{\det G^{\cdot\cdot}}}.
}
The NS5-brane remains an NS5-brane, wrapped on $\MT{3}$, and 
D6-brane becomes D3-brane. It shares with NS5 the directions
of $\MR{1,2}$. 

Now we do S-duality, so that $B^{RR}$ becomes $B^{NS}$, and NS5 becomes D5.
Also, we get the new string coupling and the new string length:
\eqn\thirdstep{
\lambda_3={l_s^3\sqrt{\det G^{\cdot\cdot}}\over\lambda},
\;\;\;\;\;
l_3=\lambda_0\sqrt{(\det g^{-1}_{\cdot\cdot})^{1\over 2}\over\rho}
}

Then, doing T-duality along the circle parameterized by $\theta$.
We have now $D6$ brane wrapped on the four-torus, and the D2 brane
inside it, orthogonal to the torus.
We end up with the following string coupling  and string length,
\eqn\finalcoupling{
\lambda_4=
{l_s^2\over\lambda_0\sqrt{\rho(\det g^{-1}_{\cdot\cdot})^{1\over 2}}},
\;\;\;\;\;
l_4=\lambda_0\sqrt{(\det g^{-1}_{\cdot\cdot})^{1\over 2}\over\rho}
}
and the following metric and $B$ field,
\eqn\metricfin{\eqalign{
ds_4^2&={l_s^2\over\rho^2}\lambda_0^2\left[ l_s^{-4}
(\det g^{-1}_{\cdot\cdot})
(d\tilde{\theta}-\epsilon_{IJK}b^{IJ}d\psi^K)^2 
+ g^{-1}_{IJ}d\psi^Id\psi^J\right],\cr
B&=\tw^I\epsilon_{IJK} d\psi^J\wedge d\psi^K+
  \etw_I d\theta\wedge d\psi^I.
}}
Let us summarize.
We have started with $k$ NS5-branes sitting at the center
of the Taub-NUT space, string coupling $\lambda_0$ and
string length $l_s$. 
The background fields are given by the equations
\metricloc\ and \bfieldloc, they correspond to both $\tw$-twists
and $\etw$-twists present. By the chain of dualities, we have mapped this
configuration to $k$ D6 branes wrapped on $\MT{4}$, and one D2 brane,
the metric and the $B$ field given by \finalcoupling\ and \metricfin.  
Notice that the volume of $\MT{4}$ is ${l_s^2\over \rho^2}l_4^4$.
In the limit we are interested in ($\lambda_0\to 0$)
it remains finite in the string units (specified by $l_4$).
The shape of the torus does not depend on $\rho$.


\subsec{World-sheet T-duality in the limit $\TNS\rightarrow \infty$}
Let us now see what happens in the limit $\TNS\rightarrow \infty$.
The strategy will be to start with type-IIA string-theory on 
the purely geometrical background which realizes the $\tw$-twist.
We will then perform world-sheet T-duality on $\MS{1}$ to obtain
a nonlinear world-sheet $\sigma$-model. Finally, we will insert the
NS5-branes back.

To describe the geometrical background we choose,
$$
X_6,\dots,X_9,
$$
as the transverse coordinates (on which the R-symmetry $SO(4)$ acts).
These replace the coordinates $y$ and $\vec{r}$ of $\TBNT(\rho)$.
We will denote,
$$
Z_1 = X_6 + i X_7,\qquad Z_2 = X_8 - i X_9.
$$
The other coordinates will be denoted,
$$
X_0\dots X_5,
$$
where $X_5$ is periodic with period $2\pi$.
They are  the world-sheet fields corresponding to
$x_0,x_1,x_2,\psi_1,\psi_2,\psi_3$ from the previous section.
The bosonic part of the world-sheet action is,
$$
L_0 = \sum_{\u,\v=0}^4 \eta^{\u\v} \px{\a}X_\u\qx{\a}X_\v
 + R^2 \px{\a}X_5\qx{\a}X_5 
+ \sum_{i=1,2} \px{\a}\bZ_i\qx{\a}Z_i.
$$
Let us, for simplicity, twist only along $X_5$ ($=\psi_3$).
The twist implies that $Z_i$ are not single-valued but rather,
$$
W_i = Z_i e^{-i{\tw\over {2\pi}} X_5},\qquad i=1,2
$$
are single-valued.
The world-sheet Lagrangian now reads,
$$
L_0 = \sum_{\u,\v=0}^4 \eta^{\u\v} \px{\a}X_\u\qx{\a}X_\v
 + R^2 \px{\a}X_5\qx{\a}X_5 
+ \sum_{j=1,2} |\px{\a}W_j + {{i\tw}\over {2\pi}}W_j\px{\a}X_5|^2.
$$
Next we perform T-duality by the standard technique of treating
$V_\a \equiv \px{\a}X_5$ as an independent field and inserting
a Lagrange multiplier $Y$ for $\px{\lbrack\a}V_{\b\rbrack}$.

The result is a world-sheet action corresponding to the metric and 
$B$-field,
\eqn\reswsac{\eqalign{
ds^2 =& \sum_{\u,\v=0}^4 \eta^{\u\v} dX_\u dX_\v
     + |dW_1|^2 + |dW_2|^2
\cr
     &+ {{dY^2 + \sum_j (i W_j d\bW_j - i\bW_j d W_j)^2}
       \over
       {R^2 + {{\tw^2}\over {4\pi^2}} (|W_1|^2 + |W_2|^2)}},
\cr
B_{\u\v}dx^\u\wdg dx^\v =&
{{dY\wdg\sum_j (i W_j d\bW_j - i\bW_j d W_j)}
       \over
       {R^2 + {{\tw^2}\over {4\pi^2}} (|W_1|^2 + |W_2|^2)}}.
}}

\subsec{Adding in the NS5-brane}
Now we repeat the same excercise with the NS5-brane metric.
In string units, the metric is,
$$
L_0 = \sum_{\u,\v=0}^4 \eta^{\u\v} \px{\a}X_\u\qx{\a}X_\v
 + R^2 \px{\a}X_5\qx{\a}X_5 
+ {1\over {|Z_1|^2 + |Z_2|^2}}\sum_{i=1,2} \px{\a}\bZ_i\qx{\a}Z_i.
$$
The dilaton is given by,
$$
g_s^2 = {1\over {|Z_1|^2 + |Z_2|^2}},
$$
and the solution is to be trusted when $g_s\ll 1$.
(See discussion in \rABS.)
After T-duality we obtain,
\eqn\reswsac{\eqalign{
ds^2 =& \sum_{\u,\v=0}^4 \eta^{\u\v} dX_\u dX_\v
     + {{|dW_1|^2 + |dW_2|^2}\over {|W_1|^2 + |W_2|^2}}
\cr
     &+ {{dY^2 + {1\over {\| W\|^4}}
                 \sum_j (i W_j d\bW_j - i\bW_j d W_j)^2}
       \over
       {R^2 + {{\tw^2}\over {4\pi^2}}}},
\cr
B_{\u\v}dx^\u\wdg dx^\v =&
{{dY\wdg\sum_j (i W_j d\bW_j - i\bW_j d W_j)}
       \over
       {(R^2 + {{\tw^2}\over {4\pi^2}})\| W\|^2 }}.\cr
}}

This is to be trusted when,
$$
\| W\|^2 \equiv |W_1|^2 + |W_2|^2  \gg 1.
$$
We see that as $R\rightarrow 0$, the $Y$-direction stays of 
finite size ${{2\pi}\over {\tw}}$.

\subsec{Large radius limit}
An interesting question is
what is the low-energy description of $S_B(k)$
compactified on $\MS{1}$ of radius $R$
with a fixed $\eta$-twist in the limit $R\rightarrow\infty$.
Naively, one can argue as follows.
To perform an $\eta$-twist we have to go over the ``fundamental''
degrees of freedom of $S_B(k)$ (whatever  they are!) and
separate them according to their charge $Q$ under the $U(1)$ subgroup
of the R-symmetry and according to their momenta $n$ 
and winding $w$ along $\MS{1}$.
We then add $\eta Q R$ to the mass of this field.
In the limit $R\rightarrow\infty$ and for generic $\eta$, this will
push all the $Q$-charged fields to high energy and we will be left with
only the $Q$-neutral sector.
Thus, if we start with $\SUSY{(1,1)}$ $U(k)$ SYM in 5+1D, as the
effective low-energy description, the conclusion would be that
we are left with $\SUSY{(1,0)}$ $U(k)$ SYM.
This conclusion cannot be correct since the gluinos of the
$\SUSY{(1,0)}$ vector-multiplet are chiral and the theory has a 
local gauge anomaly.

One possibility is that there is no 5+1D limit.
 For this to be true we must show
that there are no BPS states corresponding
to light KK states. On the type-IIA side we must show that there are no
states made 
by strings wrapped on the T-dual $\MS{1}$ which would become light.
Perhaps, when the circle is small enough, they do not form bound states
any more?


\newsec{Conclusion}

Let us summarize the results:
\item{1.}
The moduli space of the little-string theories of $k$ NS5-branes
compactified on $\MT{3}$ with $Spin(4)$ R-symmetry $\tw$-twists
is equal to the moduli space of $k$ $U(1)$ instantons on a
non-commutative $\MT{4}$.
The shape of the $\MT{4}$ is determined by the shape and
size of the physical $\MT{3}$ and by the NSNS 2-form fluxes
along it. The non-commutativity parameters are determined from the values
of the twists.

\item{2.}
In principle, there are 6 non-commutativity parameters on $\MT{4}$.
They are determined from the 3 geometrical $\tw$-twists and
the 3 non-geometrical $\etw$-twists.
The moduli space depends only on the 3 self-dual combinations
of the non-commutativity parameters and hence only on
the sum of the $\etw$-twists and $\tw$-twists.

\item{3.}
Combining the result for $k=2$ with the result of \rCGK, we obtain
a concrete prediction for the moduli space of 2 $U(1)$ instantons
on a non-commutative $\MT{4}$. This 8-dimensional
moduli space is a resolution of $(\MT{4}\times\MT{4})/\BZ_2$
by blowing up the singular locus. It can also be described as
a $\MT{4}$ fibration over a $\BZ_2^4$ quotient of a particular $K3$. 
The fiber corresponds to the ``center-of-mass'' of
the NS5-branes and the structure group is
$\BZ_2^4$ acting as translations of the fiber.
The particular point in the moduli space
of hyper-K\"ahler metrics on the $K3$ was constructed in \rCGK\
as a function of the $\tw$-twists, i.e. the non-commutativity parameters.
This $K3$ turns out to have a $\BZ_2^4$ isometry.  The $K3$ can be
described by blowing up $\MT{4}/\BZ_2$ and the $\BZ_2^4$ acts
by permuting the exceptional divisors of the blow-up. Note that
this $\BZ_2^4$ does not act freely.\foot{
In \rCGK, the global ``center-of-mass''
of the NS5-brane and the $\BZ_2^4$ were ignored,
 and only the $K3$ was studied.}

\item{4.}
Similarly,
the moduli space of the little-string theories of \rIntNEW\ of $k$
NS5-branes at an $A_{q-1}$ singularity, compactified on $\MT{3}$
with $\tw$-twists (twists in the global $U(1)$),
is equal to the moduli space of
$k$ $U(q)$ instantons on a non-commutative $\MT{4}$.

\item{5.}
We studied the phase transitions which occur at singular points
of the moduli space.

\item{6.}
If instead of the little-string theories we start with
the $(2,0)$ theory (or the SCFT theory of \rBluInt\ in item (4) above),
we obtain the moduli spaces of instantons on 
a non-commutative $\MT{3}\times\BR$.
The non-commutativity parameters are only along $\MT{3}$, which
is in accord with the fact that there are no $\etw$-twists for
this  problem.

\vskip 0.5cm
Let us conclude with 3 open problems:
\item{a.}
Generalize to other gauge groups, in particular to D-type and E-type
little-string theories.
\item{b.}
Generalize to NS5-branes at D-type or E-type singularities.
\item{c.}
Study the $\etw$-twists, in particular how they are described at
large compactification radii.

\bigbreak\bigskip\bigskip
\centerline{\bf Acknowledgments}\nobreak
We would like to thank O. Aharony, M. Berkooz, A. Schwarz and S. Sethi
for discussions.
We are also grateful to
N. Seiberg and E. Witten for discussions and helpful comments.
The work of YKC and OJG
is supported by NSF grant number PHY-9802498.
The work of MK is supported by the Danish Research Academy.
The work of AM is partly supported by RFFI Grant No. 96-02-19085
and Grant No. 96-15-96455 for support of scientific schools.
\listrefs
\bye
\end